\documentclass[preprint]{aastex}

\begin{document}

\title{Probing Fine-Scale Ionospheric Structure \\ 
with the Very Large Array Radio Telescope}

\author{A.~S.~Cohen \altaffilmark{1},
and H.~J.~A.~R{\"o}ttgering \altaffilmark{2}
}
\altaffiltext{1}{Naval Research Laboratory, Code 7213, Washington, DC, 
20375 USA, Aaron.Cohen@nrl.navy.mil}
\altaffiltext{2}{Leiden Observatory, Leiden University, Oort Gebouw, P.O. Box 9513, 2300 RA Leiden, The Netherlands}

\author{Version: \today}

\begin{abstract}

High resolution ($\sim1$~arcminute) astronomical imaging at low frequency 
($\leq150$~MHz) has only recently become practical with the development of 
new calibration algorithms for removing ionospheric distortions.  In addition
to opening a new window in observational astronomy, the process of  
calibrating the ionospheric distortions also probes ionospheric structure 
in an unprecedented way.  Here we explore one aspect of this new type of 
ionospheric measurement, the differential
refraction of celestial source pairs as a function of their angular
separation.  This measurement probes variations in the spatial gradient of 
the line-of-sight total electron content (TEC) to 
$\sim10^{-3}\,\textrm{TECU/km}$ 
(1~TECU = $10^{12}\textrm{cm}^{-2}$) accuracy 
over spatial scales of under 10~km to over 100~km.   We use data from the 
VLA Low-frequency Sky Survey \citep[VLSS;][]{cohen07}, a 
nearly complete 74~MHz survey of the entire sky visible to the Very Large 
Array (VLA) telescope in Socorro, New Mexico.  These data comprise over 
500~hours of observations, all calibrated in a standard way.
While ionospheric spatial structure varies greatly from one observation 
to the next, when analyzed over hundreds of hours, statistical patterns 
become apparent.  We present a detailed characterization of how the 
median differential refraction depends on source pair separation, elevation
and time of day.  We find that elevation effects are large, but 
geometrically predictable and can be ``removed''
analytically using a ``thin-shell'' model of the ionosphere.  We find 
significantly greater ionospheric spatial variations during the day than 
at night.  These diurnal variations appear to affect the larger angular 
scales to a greater degree indicating that they come from disturbances 
on relatively larger spatial scales (100s of km, rather than 10s of km).

\null\vskip 1in
\end{abstract}


\section{Introduction}

The degree to which the ionosphere distorts radiation increases greatly 
toward low frequencies ($\nu \leq 150$~MHz) and has been a major obstacle to 
the exploitation of this frequency range for high-resolution radio astronomy.  
Ionospheric phase variations had historically limited most low-frequency 
interferometers to relatively short baselines ($\leq$~3~km) and
consequently poor angular resolution and confusion-limited sensitivity.
An important breakthrough occurred with the development of self-calibration
\citep{pearson84} which motivated development of a 74~MHz system on the 
Very Large Array (VLA - maximum baseline 35 km) radio telescope in Socorro, 
New Mexico \citep{kassim93,kassim07}.  However, because of significant 
variations
of the ionospheric phase distortions across the field of view, self-calibration
with the 74~MHz system was limited to just a few dozen astronomical sources 
that are bright enough and isolated enough to dominate their field of view.  
It was not until the development of the first position-dependent ionospheric 
correction algorithms \citep{cotton04,cotton05} that it became 
feasible to image sources near the detection limit throughout the field of 
view.

The ability to conduct full-field imaging at 74~MHz led to the 
VLA Low-frequency Sky Survey \citep[VLSS;][]{cohen07}, a 74~MHz survey of 
the entire sky above declination $\delta > -30^\circ$.  It was taken 
primarily in 
the VLA B-configuration which has maximum baseline lengths of about 11~km.  
The primary motivations for the VLSS were astronomical, and indeed 
ionospheric effects were a contamination to be removed from the 
celestial images of interest.  Nevertheless, in the process of 
removing these ``contaminations'', ionospheric measurements were produced.  
The $>$500 hours of observations taken in the B-configuration for this 
survey constitute an unprecedented data set with which to characterize the
ionosphere.  This paper presents an initial look at the fine-scale ionospheric
structure seen in these data with a statistical study of the differential 
refraction of simultaneously imaged pairs of sources.  We explore what this 
means both for ionospheric science itself as well as for
predicting the nature and degree of ionospheric distortions
that will be faced by a new generation of planned low-frequency telescopes
such as 
LOFAR \footnote{http://www.lofar.org}, 
LWA \footnote{http://lwa.unm.edu}, 
MWA \footnote{http://www.haystack.mit.edu/mwa}
and the upcoming 50~MHz system on the 
GMRT \footnote{http://gmrt.ncra.tifr.res.in}.

Previous studies of the ionosphere with low-frequency radio interferometers
have concentrated on using measured phases from individual antennas to detect 
traveling ionospheric disturbances, most likely due to gravity waves 
(see \citet{jacobson92} and references within).  This study differs from 
these previous studies in several important ways.  First, the previous 
studies used 
Fourier analysis of the phases of individual antennas, and therefore were 
mainly sensitive to traveling sinusoidal ionospheric disturbances that were 
persistent in time.  Our study uses no initial model of the ionospheric 
structure, and is sensitive to wave and non-wavelike structure, such as 
turbulence.  Second, because these studies only measured single 
sources, the region of the ionosphere probed was limited to the size of 
the array (all lines of sight from antennas to that one source) which was 
35~km for the \citet{jacobson92} study (the VLA in A-configuration).  
Because our study used the entire field of view, it probed a region of 
the ionosphere that can be well over 100~km depending on the elevation of the 
field.  Finally, while previous studies relied on one or a few several-hour
observations, this study is based on over 500 hours of observations taken 
over the course of several years.  Thus, rather than obtaining ``snapshot''
views of ionospheric behavior, we were able to create an overall statistical
description of typical ionospheric structure and how this depends on various 
factors, such as elevation and time of day.

This paper is organized as follows.  Section~\ref{physics.sec} quantitatively 
describes the impact of the ionosphere on incoming radiation at radio
frequencies, how this affects data obtained from an interferometric array, 
and the specific measurements that we have obtained.  In Section~\ref{dr.sec}
we present and analyze the differential refraction data obtained from the 
VLSS and its dependence on elevation and time of day.  In  
Section~\ref{conclusions.sec}, we discuss what conclusions can be made from 
the analysis of the differential refraction statistics.  Finally, in 
Section~\ref{future.sec} we describe the many possibilities for future 
ionospheric studies based on these types of measurements.

\section{Radio Interferometric Observations through the Ionosphere}
\label{physics.sec}

\subsection{Ionospheric Refraction}

The ionosphere causes a phase delay to incoming radio waves which is that 
for radiation passing through a plasma and can be approximated by:
\begin{equation}
\phi \approx \frac{e^2}{c\,m_e\nu}\int{N_e(l)\,dl}
\label{delay1.eqn}
\end{equation}
where $\phi$ is the phase delay in radians, $e$ is the electron charge, 
$m_e$ is the electron mass, $c$ is the speed
of light in a vacuum, $\nu$ is the 
light frequency and the integral is the column density of free electrons
($N_e(l)$) along the line of sight \citep{rybicki79}.  For a source at 
zenith and referencing to the VLSS frequency of $\nu =$ 74~MHz, this 
can be simplified to:
\begin{equation}
\phi = 1.14\times10^2\,\left(\frac{\nu}{74\,\textrm{MHz}}\right)^{-1}\,
\left(\frac{\textrm{TEC}}{1\,\textrm{TECU}}\right)
\label{TEC.eqn}
\end{equation}
where TEC stands for total electron content and a TECU (or TEC unit) is a unit 
of vertical electron column density equal to $10^{12}$ electrons/cm$^2$.

In the limit of small refraction angles (total refraction $\leq 10^{-3}$ 
radians), the refracted path passes 
through very nearly the same ionosphere as the un-refracted path, and so the 
total phase delay of a single ray can be approximated by integrating the 
total ionospheric delay along the original straight-line path.  Therefore, 
a proper ray tracing is unnecessary.  We follow this ``small refraction 
angle'' approximation throughout this paper.

\begin{figure}
\plotone{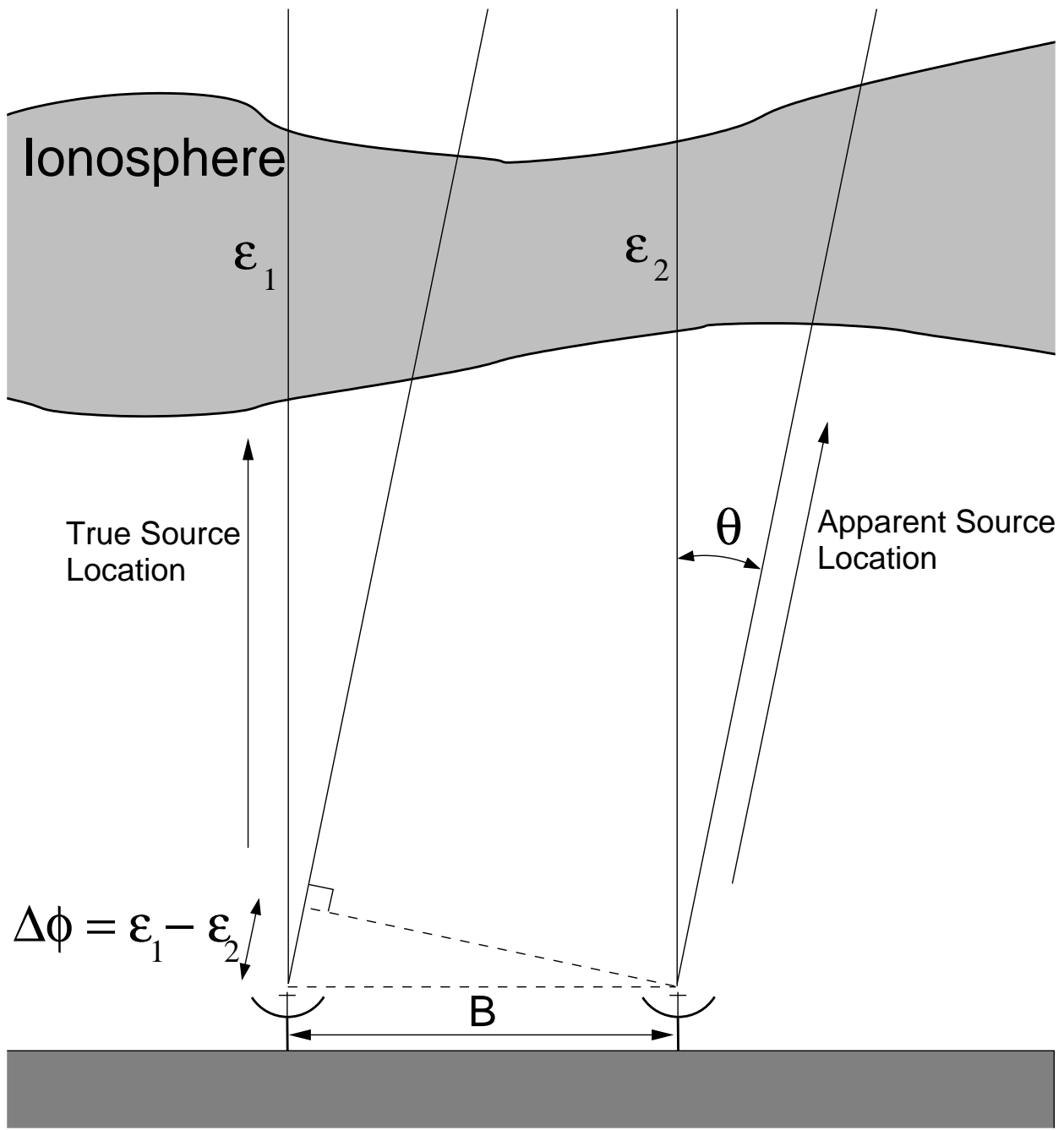}
\caption{Different phase delays, $\epsilon_1$ and $\epsilon_2$, between 
antennas are interpreted by an interferometer as a geometric phase delay, 
$\Delta\,\phi = \epsilon_1 - \epsilon_2$, due to the incoming signal 
emanating from a different direction.  This assumes that the refraction angle, 
$\theta$, is small enough that the refracted paths pass through roughly the 
same ionosphere as the non-refracted paths and so ray tracing is not 
necessary.
\label{baseline.fig}}
\end{figure}

As a radio interferometer only 
measures {\em differences} between signals received at pairs of antennas, 
a constant phase delay across an interferometric array will have no effect.
If the phase delay is different from one antenna to another, the phase 
difference between the antennas produces a geometric phase delay due to 
the wave front coming from a different direction as seen in 
Figure~\ref{baseline.fig}.  This phase delay between the paths arriving 
at two antennas causes an apparent position shift in the source location 
by an angle:
\begin{equation}
\theta = \frac{\lambda\,\Delta\phi}{2\pi B}
\label{shift.eqn}
\end{equation}
where $\lambda$ is the wavelength of the light ray, $B$ is the baseline 
length, and $\Delta\phi$ is the difference in ionospheric phase delays 
between the two antennas.  If the ionospheric phase delay varies linearly
across an array of antennas, all baselines ``see'' the same source shift, 
and the only effect on the resulting image of a celestial source is that 
it appears shifted from its true sky position.  Combining 
Equations~\ref{TEC.eqn} and \ref{shift.eqn}, we can relate the gradient of 
the TEC along a direction $x$ on the ground to the observed angular shift
in that direction, $\theta_x$, by
\begin{equation}
\frac{d}{dx}\textrm{TEC} = 
6.59\times10^{-5}\,\textrm{TECU/km}\,
\left(\frac{\nu}{74\,\textrm{MHz}}\right)^2\,\left(\frac{\theta_x}{1''}\right).
\label{gradient.eqn}
\end{equation}
Note the strong dependence on frequency, $\nu$, 
which explains why the ionosphere is mainly a problem for low-frequency
radio observing.
If the ionospheric phase delay does {\em not} vary
linearly across the array, but also has higher order terms (curvature), 
the image is not only shifted, but also distorted.
However, in the regime of 74~MHz VLA observations with the B-configuration 
(11~km array), this distortion is rarely significant except under unusual 
ionospheric conditions.  

\subsection{Differential Refraction}

In practice, absolute positions of sources are difficult to measure 
because that would require differentiating between the ionospheric phases
and instrumental phases during the initial calibration of the array.  The 
instrumental phases are determined by observing a bright calibrator
source, but these solutions are degenerate to a phase gradient across the 
array corresponding to the unknown ionospheric position offset of that 
calibrator source.  That phase gradient does not affect astronomical 
observations as it is corrected in later calibration steps.  But it does
corrupt the measurement of ionospheric position offsets of other observed 
sources.  Unless this effect is determined and removed, which has yet to be
done, it corrupts the measurement of absolute ionospheric gradients.  

What can be measured with great accuracy is the change of ionospheric phase
gradients across a single field of view.  While the absolute ionospheric 
offsets of sources in this field of view are corrupted by an arbitrary phase 
gradient applied in the initial calibration, 
the {\em differences} in the position shifts of two different sources in 
the image can be determined very accurately because the initial corruption
applies equally to both sources and so cancels out in the difference.  
From now on we will refer to these differences in position shifts as the
``differential refraction'' between pairs of sources measured simultaneously
in the same field of view.  Source pair differential refraction measures 
the difference in the ionospheric phase gradient between the paths leading 
to the two source locations.  

\subsection{What Differential Refraction Measures}

\begin{figure}
\plotone{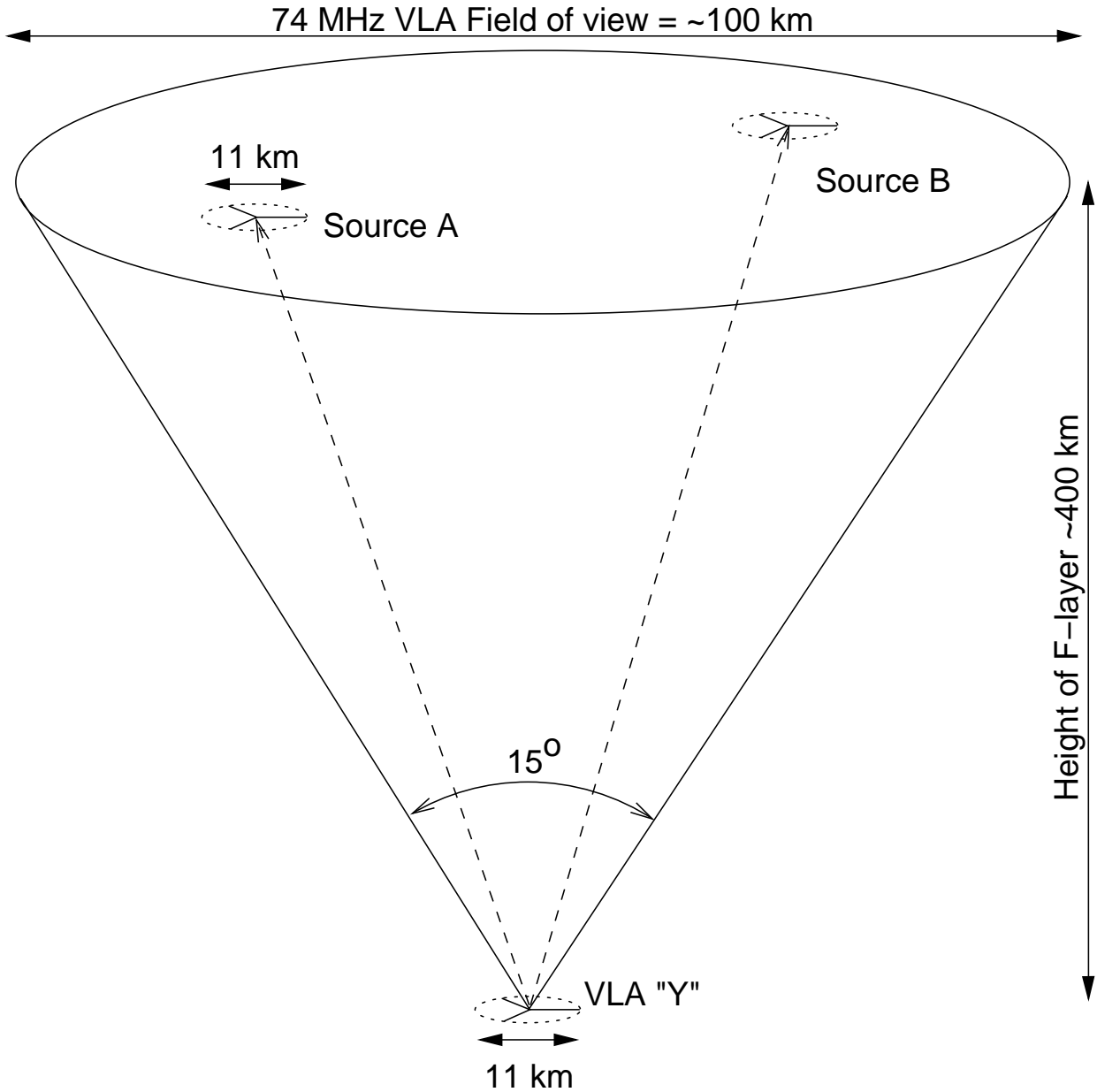}
\caption{Relative sizes of the VLA in B-configuration to the field of view
at 74~MHz projected to the ionospheric F-layer at a height of 400~km.  In the
direction of a source ``A'', the relative phases through the ionosphere are 
measured along the 11~km projection of the VLA ``Y''.  For another source ``B''
in the field of view, the phases in a different 11~km region are measured.  
Within each projected region, the phases are assumed to vary linearly, though
that phase gradient may be different for the two sources.  The differential 
refraction measures the phase gradient difference between the two source 
locations, which probes the phase screen curvature over the field of view.
All sizes in this diagram are to scale except for the vertical direction which
is compressed for easier viewing.
\label{patches.fig}}  
\end{figure}

At this point it is useful to relate differential refraction to the 
geometry of the ionosphere.
The diffraction-limited field of view at 74~MHz for the 25-meter
VLA antenna dishes is about $15^\circ$ across.  While this field of view 
probes a cone-shaped region through the ionosphere, most of the electrons are 
usually located within a relatively thin layer near the altitude of maximum 
electron density in the F-layer \citep{kelley89,schunk00}.  It is therefore 
instructive to consider a thin-shell 
model of the ionosphere, in which the ionosphere is approximated as a very 
thin layer at a constant height above the Earth.  
For a shell height of 400~km (see Section~\ref{elevation.sec}),  
the field of view cone intersects the shell in a circle of diameter 
$\sim$100~km at zenith.  In the 11~km B-configuration, the lines of sight 
from individual VLA antennas toward any single source define a set of 
pierce points in an 11~km region corresponding to the VLA projected to the 
F-layer 
altitude (Figure~\ref{patches.fig}).  The observed position offset of that
source is determined by the average ionospheric phase gradient in that set 
of pierce points.  The assumption that for individual sources ionospheric
distortions are primarily refractive is equivalent to assuming that in any 
11~km region of pierce points, the ionospheric variations can be 
approximated as linear.  However, this refraction, or phase gradient, may 
be different at the location 
of another 11~km region towards a different source in that field of view, and 
thus the refraction across the field of view changes and the image becomes 
``warped''.  Differential refraction measures the magnitude of the vector 
difference in ionospheric phase gradients between the ionospheric pierce 
points towards two celestial sources.  

Measuring differential refraction rather than absolute refraction may seem 
like a major limitation.  However, the differential refraction of source 
pairs is actually more relevant to the unique problems of high-resolution
low-frequency radio imaging than the absolute refraction of an image.  This
is because any constant refraction across the field of view causes nothing 
more than a constant position shift of the entire image, which is easy to 
correct and remove using existing calibration techniques 
(such as self-calibration) commonly used at 
higher frequencies ($\geq$1 GHz).  The {\em differences} in the ionospheric 
phase gradient cause the refraction to vary throughout the field of view 
causing a stretching or ``warping'' of the image which standard 
self-calibration cannot correct.  Additionally, this 
warping is time-variable on a timescale of minutes or less at sub-arcminute 
resolution.  It is this warping of the field that 
will be the major challenge for future low-frequency instruments to handle, 
and differential refraction measurements probe this effect directly.

As for studying the ionosphere, the absolute refraction corresponds to an
overall TEC gradient across the entire $\sim$100~km field of view.  
Differential refraction measures higher order structure in addition to that 
overall gradient.  To be limited to differential refraction is 
to be sensitive only to finer-scale structural variations than 
the roughly 100~km field of view.  Larger scale TEC variations can be 
explored by current techniques such as line-of-sight TEC measurements 
from ground or space-based GPS receivers, UV limb scans of recombination
emission, radar and ionosonde.  Radio interferometers therefore can extend the 
measurement of ionospheric structure to much smaller spatial scales than 
are currently possible.

\section{Differential Refraction Statistics}
\label{dr.sec}

\subsection{The Data}

Ionospheric calibration for the VLSS was done by measuring the relative 
position shifts of all sources in the field of view that are detected 
within a 1-2 minute time interval.  That time interval has been determined
empirically as the maximum length of time during which time variations in 
the source shifts are small compared to the synthesized beam.  The synthesized
beam for the VLSS has a full-width at half maximum of 80$''$ 
($3.9\times10^{-4}$ radians) which corresponds to a physical size of about 160 meters
at a zenith height of 400 km.
A two-dimensional ``phase screen'' of ionospheric variations is fit to
the position shifts of the detected sources for each time interval.
The corrections are applied in the image plane.  This method is called 
``field-based'' calibration 
\citep{cotton04,cotton05} and its specific implementation for the VLSS is 
described in detail in \citet{cohen07}.  
This calibration method works remarkably well for the 74~MHz VLA in most 
circumstances, producing high quality images that are astrometrically 
accurate within about 1/4 of a synthesized beam-width and with a typical 
ionospheric smearing that averages only about 1/10 of a synthesized 
beam-width.  For some fraction of the time, the ionosphere was so turbulent
that either it was impossible to image the the bright calibrator sources
or it was impossible to fit a 2nd order Zernike polynomial to the source
offsets.  This was typically 10-20\% of the total observing time.
 
The data set used for this ionospheric study consists of the position offset 
measurements for bright ($\geq 3~Jy$\footnote{A Jy (Jansky) refers to the 
unit of flux density such that 1 Jansky = $10^{-26}\,W\,Hz^{-1}\,sr^{-1}$.}) and 
compact sources in the field of 
view that were taken and recorded during the field-based ionospheric 
calibration of all VLSS data acquired in the 11~km B-configuration.  This 
resulted in a list of source position offsets for roughly 3-8 sources for 
each 2-minute interval in over 500~hours of observing time.  In 
addition to position offsets, the fitted peak brightnesses and integrated 
flux densities 
for each calibrator source were also recorded.  The time of day was also 
recorded for each time interval, and from this and the source positions the 
elevation above the horizon was calculated.  

Some radio sources are bright enough that their side-lobes can contaminate the
detection of calibrator sources.  For this reason, we did not use field
pointings within 15$^\circ$ of Cassiopeia A or Cygnus A, or within 8$^\circ$ 
of Hydra A, the Crab, Virgo A or Hercules A.  
In order to remove false source detections and improve the accuracy of the 
position offset measurements used, we restricted our study to sources within 
$9^\circ$ of the field center, with fitted peak brightnesses above 
4~Jy/beam\footnote{Units of Jy/beam refer to Janskys (1 Jansky = $10^{-26}\,W\,Hz^{-1}\,sr^{-1}$) per synthesized beam.} that were less than 300$''$ from their known locations.  
After applying these restrictions, there were 16,032 usable 2-minute time 
intervals in the survey.  During these time intervals there were 96,575 source 
detections, and 299,359 pairs of simultaneous source detections.

We can estimate the position error along each axis by using the results of 
\citet{condon97} applied to the case where we have (1) a point source, (2) 
a circular synthesized beam, and (3) the full-width half maximum of the 
Gaussian noise correlation function equal to that of the sythesized beam.
This results in a position error estimate of:

\begin{equation}
\sigma_{pos} = \frac{\theta_b}{r\,\sqrt{8\,ln(2)}}
\label{pos.err.eqn}
\end{equation}

where $r$ is the signal-to-noise ratio of the detection, and $\theta_b$ 
is the full-width half maximum of the synthesized beam.  Because the 
average RMS noise level in the 2-minute snapshot images is about 
$\sim$0.6~Jy/beam, the 4~Jy/beam brightness cutoff ensures that we used 
sources detected with signal-to-noise ratios of 6.67 or better.  Applying 
Equation~\ref{pos.err.eqn} then indicates that the predicted 
position errors are $5.1''$ or less.

\subsection{Characterizing the Differential Refraction}

For each pair of simultaneous source detections we calculated the magnitude
of their differential refraction as:
\begin{equation}
r_{ij} = |\vec{r}_{ij}| = |\vec{r}_i - \vec{r}_j|
\end{equation}
where $\vec{r}_i$ and $\vec{r}_j$ are the measured vector position offsets of 
sources $i$ and $j$ with respect to their known positions taken from the 
higher frequency (1.4 GHz) Northern VLA Sky Survey \citep[NVSS;][]{nvssref}
and $\vec{r}_{ij}$ is the vector difference between $\vec{r}_i$ and 
$\vec{r}_j$.  If the two components of the vector $\vec{r}_{ij}$ were 
independent and normally 
(Gaussian) distributed with RMS of $\sigma$, the magnitude, $r_{ij}$, would 
have a Rayleigh distribution described by the following probability density 
function: 
\begin{equation}
P(r_{ij}|\sigma) = \frac{r_{ij}}{\sigma^2}\exp\left(\frac{-r_{ij}^2}{2\sigma^2}\right).
\label{rayleigh.eqn}
\end{equation}

\begin{figure}
\plotone{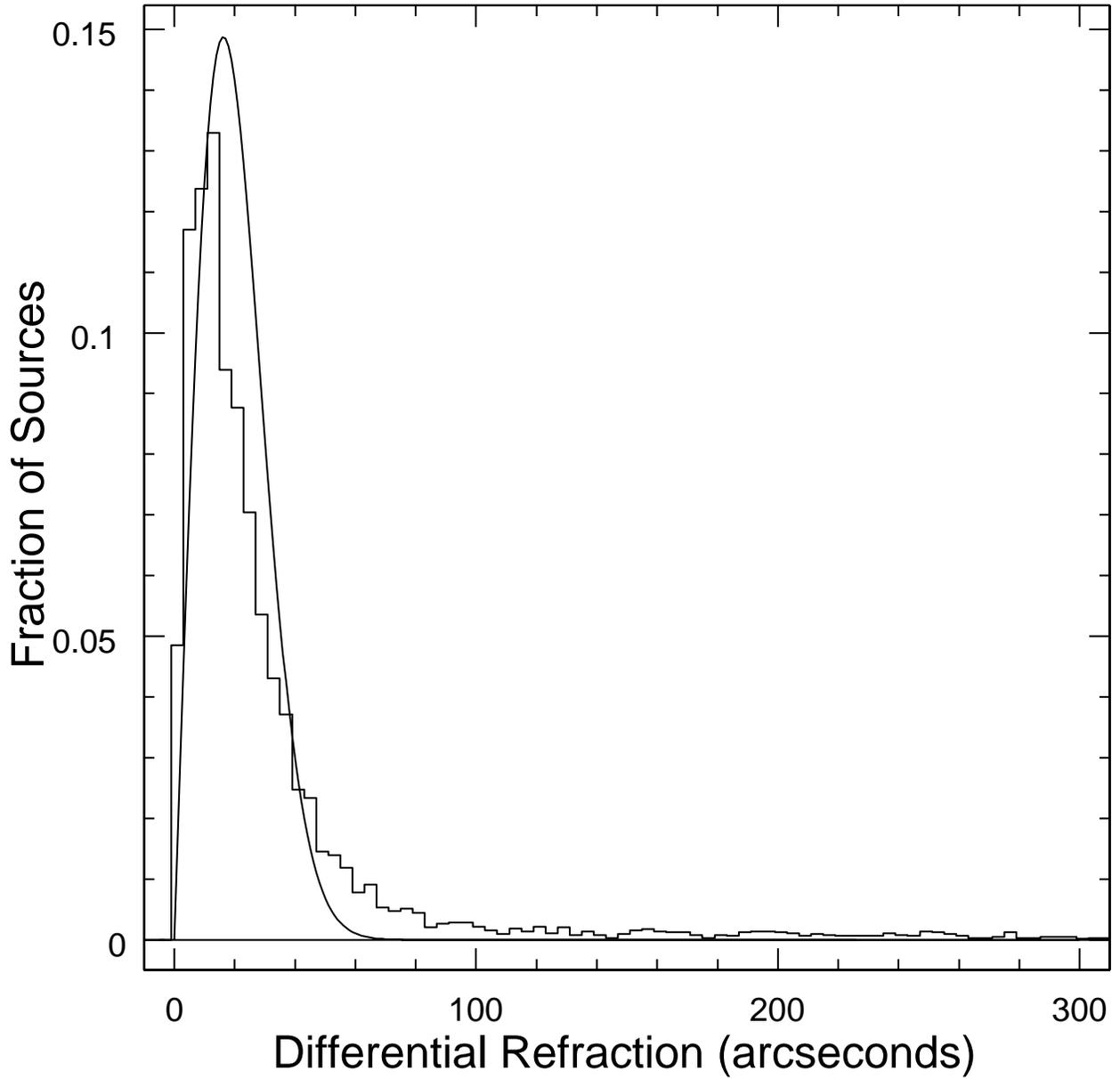}
\caption{Histogram of differential refraction for all 6,387 source 
pairs that were separated by between $5^\circ$ and $5.5^\circ$ degrees, and 
above an elevation of $60^\circ$ above the horizon.  The best-fitting 
Rayleigh distribution (smooth solid line) is clearly a poor description 
of the actual distribution which shows relatively more data points at higher
values than in the main peak.  This causes the mean to be dominated by a 
small number of data points with very strong ionospheric distortions, and 
therefore be far higher than the median.  For
this reason, the median values of the differential refraction were used, rather
than the mean values.  In this case the median is 19.2$''$ while the mean 
is 33.1$''$.  The synthesized beam has a FWHM of 80$''$.\label{ds.hist.fig}}
\end{figure}

To measure the actual probability distribution of $r_{ij}$, we needed to 
avoid variations in $r_{ij}$ caused by source pair angular separation and 
elevation.  Therefore, we considered source pairs with angular separations 
in the narrow range between $5^\circ$ and $5.5^\circ$ and only elevations 
higher than $60^\circ$ above the horizon, where elevation effects should be
small.  There were 6,387 such pairs of sources, and their distribution of 
$r_{ij}$ is shown as the histogram in Figure~\ref{ds.hist.fig}.  Also plotted
in Figure~\ref{ds.hist.fig} as the smooth solid line is the best fitting 
Rayleigh distribution.  Clearly, the Rayleigh distribution is a poor fit, 
mostly because of the relatively greater number of very high values of 
$r_{ij}$ which extend into a very long ``tail'' which persists at a 
non-negligible level for well past 10 times the peak of the distribution.  
(This, of course also results in a corresponding deficit of values within 
the main peak of the distribution.) 
This indicates that the components of the two-dimensional
vector $\vec{r}_{ij}$ are not normally (Gaussian) distributed, but rather have
a statistical overabundance of very high values, presumably corresponding 
to times of unusually disturbed ionospheric conditions.  These very high 
values of $r_{ij}$ constitute a small fraction of the data and so have 
relatively little effect on the median value of $r_{ij}$.  However, because 
those values of $r_{ij}$ are so high, they have a more significant effect 
on the mean value of $r_{ij}$.  Indeed the mean value of $r_{ij}$ for the 
distribution in Figure~\ref{ds.hist.fig} is $33.1''$, which is a full 72\% 
higher than the median value of $19.2''$.  For a Rayleigh distribution the 
mean and median are $\sqrt{\pi/2}\,\sigma$ and $\sqrt{\ln{4}}\,\sigma$ 
respectively, which 
differ by only 6\%.  Our goal is to measure {\em typical} ionospheric 
conditions.  Therefore the median value of $r_{ij}$, which is less affected
by a small number of extreme events, is the more robust measurement, 
and that is what we use for all further analysis.  We define the term 
$D(\theta_{ij})$ to be the median magnitude ($r_{ij}$) of differential 
refraction for sources with angular separation of $\theta_{ij}$.

There is no analytical formula for the standard deviation in the median 
value of a non-standard distribution.  However, if we assume that the 
standard deviation is proportional to $N^{-1/2}$ (where $N$ is the number 
of source pairs) and to the median differential shift we have:
\begin{equation}
\label{error.eqn}
\sigma (D(\theta_{ij})) = A\,\frac{D(\theta_{ij})}{\sqrt{N}}
\end{equation}
where $A$ is a constant of proportionality.  We can determine $A$ by further
examining the sample in the previous paragraph.
We can estimate $\sigma (D(\theta_{ij}))$ for this case by taking random 
samples of $N$ = 1000 source pairs from the overall sample 
of 6,387 source pairs and measuring the variance in the resulting medians.  
Taking 10,000 such sub-samples, we get a RMS variation in the median 
differential refraction of $\sigma (D(\theta_{ij})) = 0.76''$.  
We already know that $D(\theta_{ij}) = 19.2''$, and can solve for the 
remaining variable, $A$ = 1.25.

\begin{figure}
\plotone{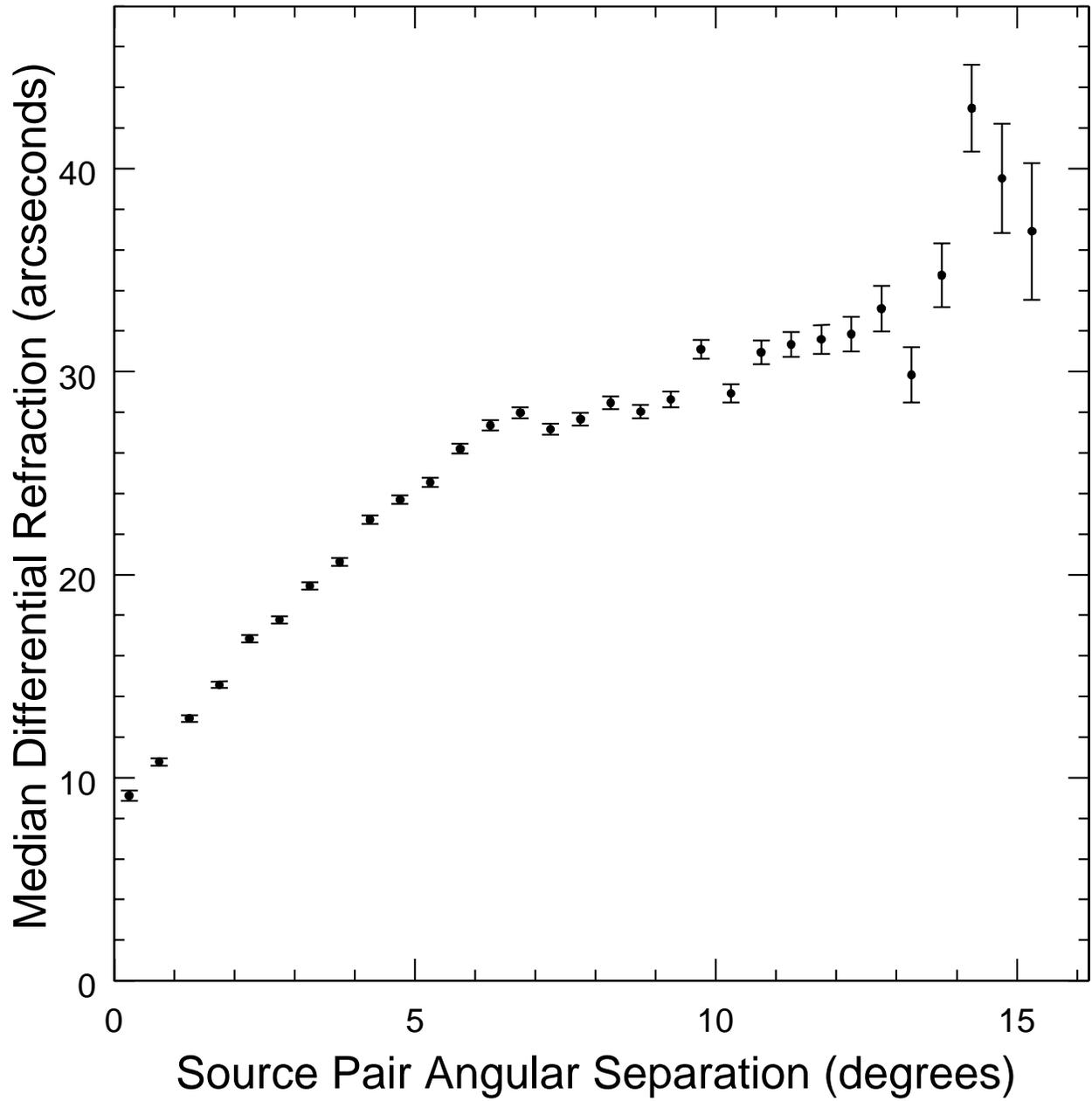}
\caption{Median differential source shifts as a function of the angular 
separation.\label{ds.noadjust.fig}}
\end{figure}

\begin{figure}
\plotone{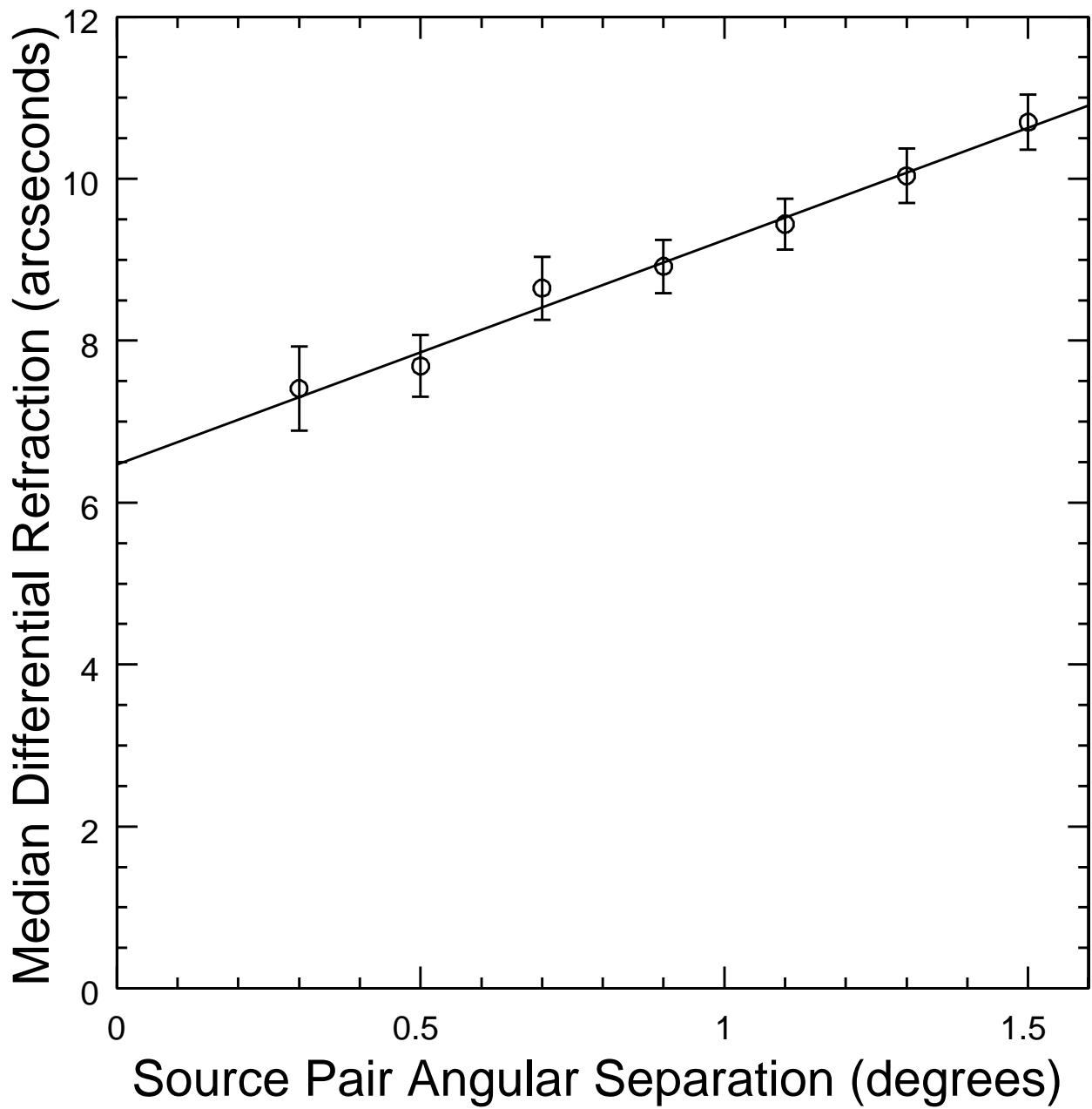}
\caption{Median differential source shifts as a function of the angular 
separation for source pairs with small angular separations and elevations 
greater than 60$^\circ$ above the horizon.  The solid line is a linear 
least-squares fit to the data points, and intersects the y-axis at a 
median differential refraction of $6.467''$.
\label{ds.close.fig}}
\end{figure}

We estimate $D(\theta_{ij})$ itself by binning the source 
pairs according to angular separation, taking the median value of the 
magnitude of the differential shifts.  The errors are then calculated 
according to Equation \ref{error.eqn}.  The resulting function is shown in 
Figure~\ref{ds.noadjust.fig}.  Because of the field-of-view limitations, 
there are relatively fewer pairs of points with large angular separations 
and so the error-bars are larger for the largest separations.  There were 
pairs with separations of more than $16^\circ$, but not enough for meaningful
statistics, so we cut off the plot at this value.  Note that the 
median differential source shift generally increases with source separation, 
though the increase is slower at larger separations.  Also note that the 
differential refraction in the limit of zero angular separation, 
$D(\theta_{ij} \rightarrow 0)$, does not appear to be zero.
This can be seen more clearly in 
Figure~\ref{ds.close.fig} which plots the median differential refraction 
for small angular separations with smaller bins so that the trend towards
zero angular spacings can be seen more clearly.  Figure~\ref{ds.close.fig}
only shows source pairs that are at high elevation ($\geq~60^\circ$) to remove
most elevation effects.  A linear least squares fit of the relation between 
angular separation and median differential refraction estimates by 
extrapolation a value of $D(\theta_{ij} \rightarrow 0) = 6.467''$.

There are two possible reasons why $D(\theta_{ij} \rightarrow 0)$ is 
non-zero.  First is that 
there are significant variations in ionospheric spatial gradients on an 
angular scale well below about 0.3$^\circ$.  Such an angular scale would 
correspond to a spatial separation of about 2~km at a characteristic 
ionospheric height of 400~km (see Section~\ref{elevation.sec}).  But that is 
much smaller than the 11~km size of the VLA in B-configuration, and therefore
would violate our assumption of ionospheric variations being roughly 
linear within the array size.  If that were the case, sources would typically
be significantly distorted rather than simply being refracted.  This occurs
in unusually bad conditions, but is not seen typically.

The second, and more likely, explanation for a non-zero value of 
$D(\theta_{ij} \rightarrow 0)$
is map-noise induced errors in the position measurements of 
the sources.  Thus, rather than tending toward zero, 
$D(\theta_{ij} \rightarrow 0)$ tends to the measurement error.  The 
measurement error can be estimated again with Equation~\ref{pos.err.eqn}.
The median peak brightness for a source in our sample is 
6.88~Jy/beam.  The typical noise level for the two-minute snapshot observations
used to measure source positions is about 0.6~Jy/beam, for a signal-to-noise ratio of $\rho = 11.5$.  For a resolution of 
$80''$, this implies a measurement error of 2.96$''$ for each axis.  A 
differential refraction measurement consists of four such measurements 
(two axes each for two sources) and so the combined error 
(added in quadrature) is 5.92$''$.  This estimate is less than 10\% lower 
than what the data show, and so this explanation seems reasonable.  Ultimately
we are interested in the median differential refraction caused by the 
ionosphere, not measurement errors.
Therefore, for all future analysis, we will subtract the value of $6.467''$
in quadrature from all median differential refraction estimates.

\subsection{Dependence on Elevation}
\label{elevation.sec}

\begin{figure}
\plotone{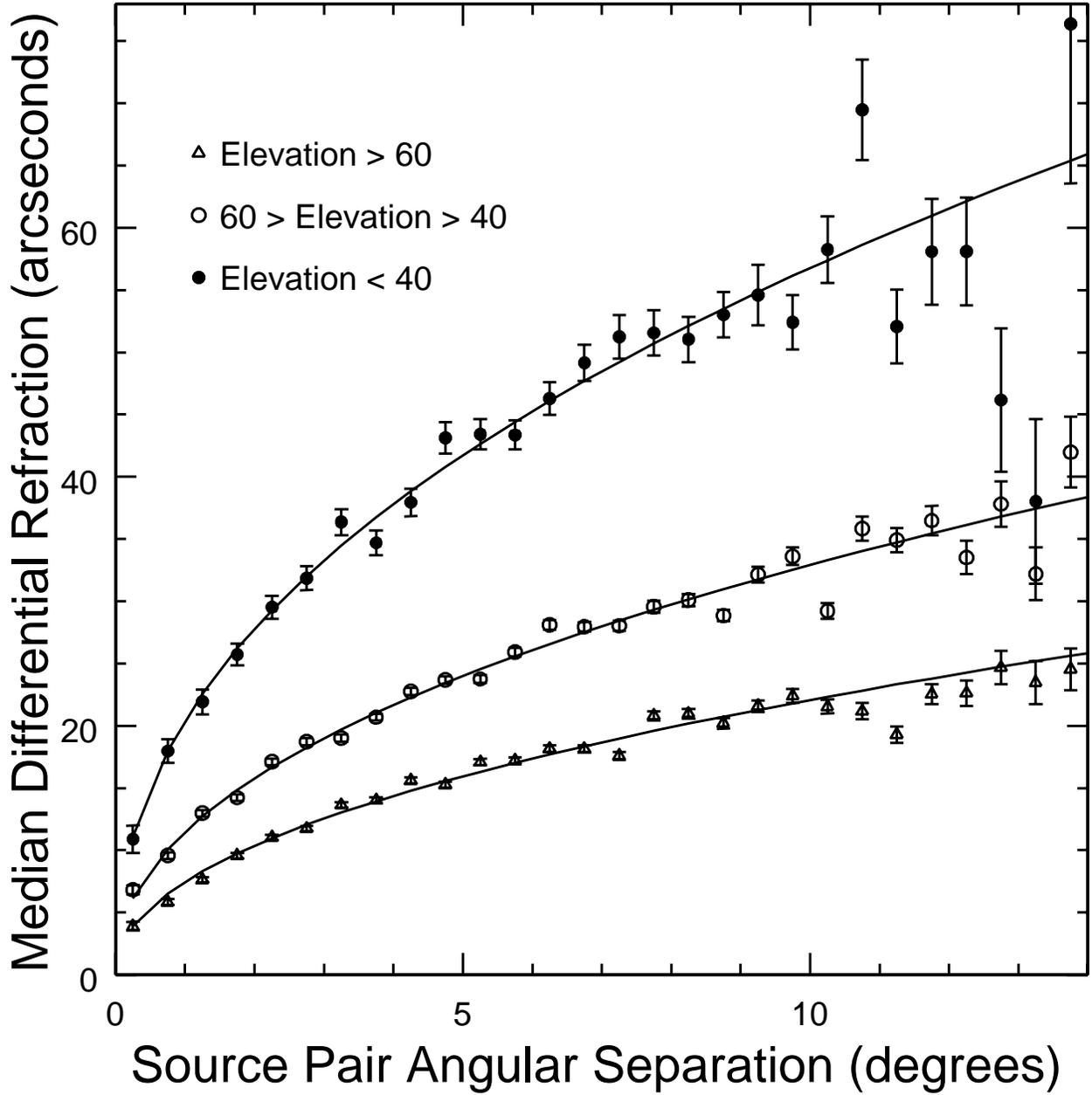}
\caption{Median differential refraction as a function of source pair separation
for data from three elevation bins.  The solid lines show power law fits to
each data set, the parameters of which are listed in Table~\ref{elev.fit.tab}.
\label{ds.raw.elev.fig}}
\end{figure}

We examined the dependence of differential refraction on elevation by 
splitting the data into three elevation bins as shown in 
Figure~\ref{ds.raw.elev.fig}.  For each elevation bin, we fit the data to 
a power-law function of the form:
\begin{equation}
D(\theta) = \beta\,\left(\frac{\theta}{1^\circ}\right)^\alpha
\label{powerlaw.eqn}
\end{equation}
where $\theta$ is the source-pair angular separation and $D(\theta)$
is the median differential refraction within a bin centered on $\theta$.  
The exponent, $\alpha$,
and scale factor, $\beta$, were determined by fitting a linear model to the 
logarithms of the data points using the least squares method.  
The resulting power-law fits are shown as the solid lines in 
Figure~\ref{ds.raw.elev.fig} and their parameters are listed in 
Table~\ref{elev.fit.tab}.  As Figure~\ref{ds.raw.elev.fig} demonstrates, 
the data are fairly well described by a power law fit.  It is also clear that 
the differential refraction increases significantly as elevation decreases.

Because elevation is mainly a geometrical effect, it should be possible
to analytically ``remove'' its effects on the data. We do this 
with the commonly used ``thin-shell'' model of the ionosphere in which 
the ionosphere is approximated as very thin layer encircling the Earth at 
a constant height above the Earth's surface \citep{lanyi88,ma03}.  In 
this model, the entire ionospheric phase delay of incoming radiation occurs 
at the ``pierce-point'' where the observing path to the celestial source 
intersects the ionospheric shell.  
Generally, the shell-height above the Earth's surface, $h$, is taken
to be the height of maximum electron density, with values between about 
300 to 500~km typically used.  In particular, \citet{nava07} found that a 
shell height of $h = 400$~km minimized errors when using the thin-shell model
to convert from slant to vertical GPS TEC measurements, and we use that 
value for all upcoming calculations.  

According to this formulation, elevations affects the measurement of 
differential refraction in two ways.  First, it changes the relationship 
between the angular separation of a source pair and the spatial separation 
of the ionospheric pierce-points of the observing paths to those sources.
We correct the observed source-pair angular separation for this by 
calculating the spatial pierce-point separations for 
each source pair based on the altitude and azimuth of each source, and
converting this to the equivalent angular separation if the sources 
were centered at zenith.  Second, elevation determines the effective 
path length through the ionosphere.  For an angle of incidence, $\epsilon$ 
(where $\epsilon = 90^\circ$ at zenith), the effective path length through
the ionosphere is increased by a factor of sec($\epsilon$).  The 
ionospheric phase delays also increase by this factor 
(see Equation~\ref{delay1.eqn}), as does the effective 
gradient of the phase delays.  Because sources within each pair are a few
degrees apart, they have slightly different values of sec($\epsilon$), 
and we used the quadratic average to convert the measured differential
refraction to its predicted zenith value.

\begin{figure}
\plotone{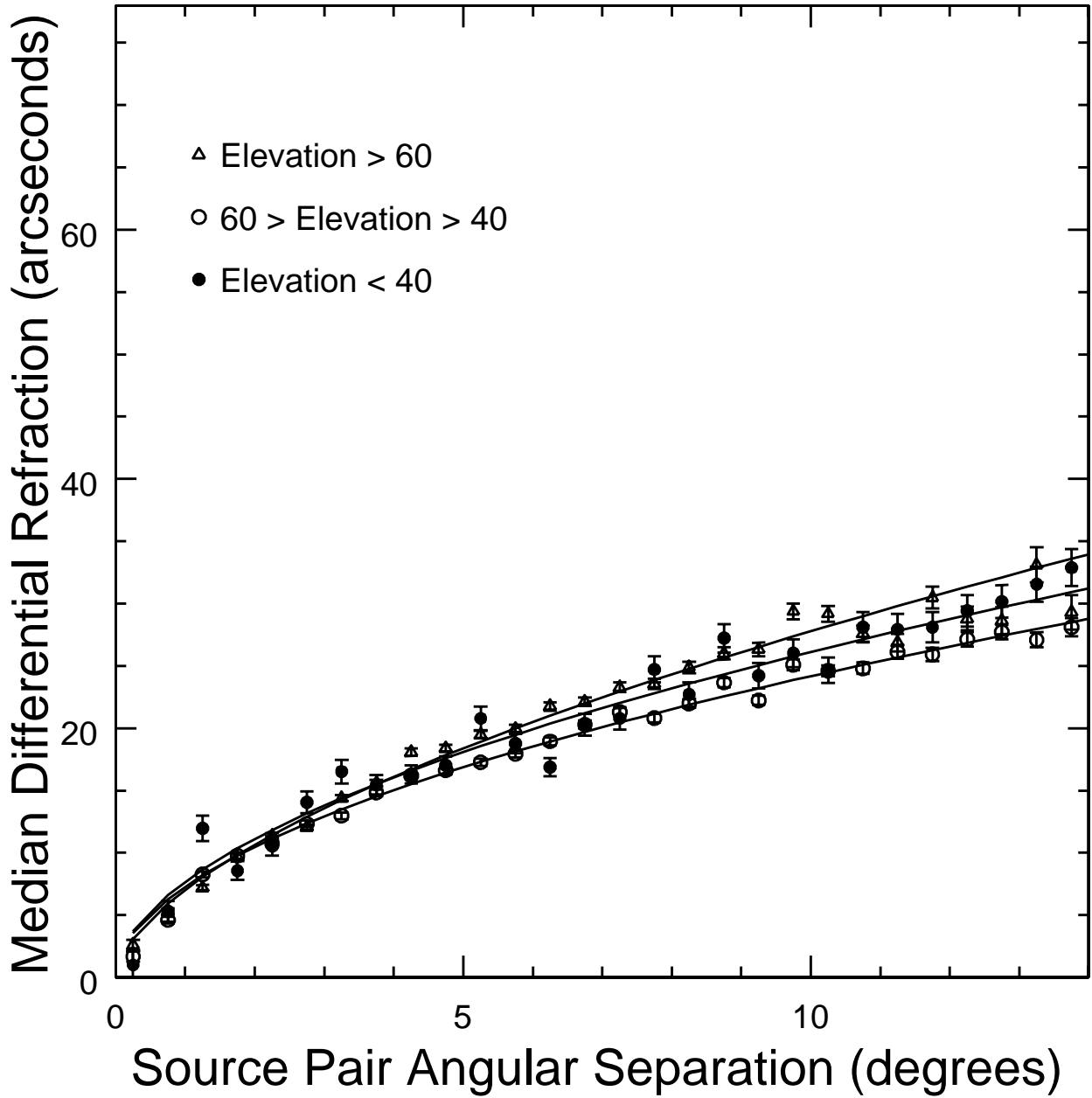}
\caption{Median differential refraction as a function of source pair separation
for data from three elevation bins for data converted to equivalent values
at zenith.
\label{ds.elev.fig}}
\end{figure}

\begin{figure}
\plotone{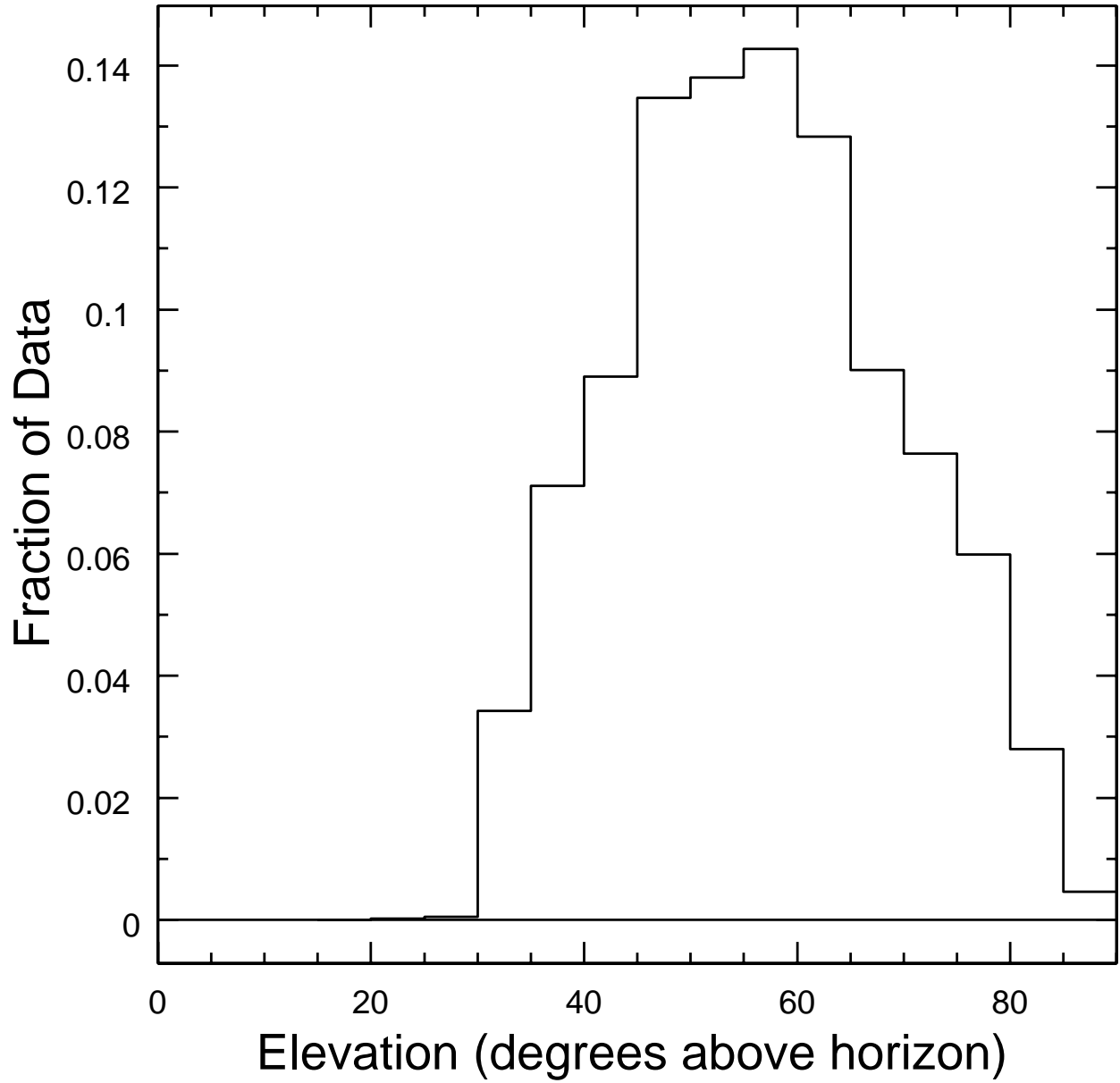}
\caption{Histogram of the fraction of data taken as a function of elevation above the horizon.
\label{Elev.hist.fig}}
\end{figure}

These two adjustments attempt to convert all differential refraction and
angular separation measurements to what they would be if the 
measurement had been taken for source pairs centered at the zenith.  
Figure~\ref{ds.elev.fig} shows the same data as Figure~\ref{ds.raw.elev.fig},
but with the differential refractions converted to their zenith equivalent
as described above.  As can be seen by the data points and 
the fitted power law curves, the applied elevation adjustments have 
removed nearly all differences between the three elevation bins.  
We note that our data cover a wide range of elevations, but generally 
avoid elevations below $30^\circ$ (see Figure~\ref{Elev.hist.fig})  
\footnote{This is because it was already known anecdotally that the ionospheric distortions are greater at low elevations, and the original goal of the observations was astronomical rather than ionospheric.}. 
So it is unclear if these corrections continue to work at lower elevations 
however, as Figure~\ref{ds.elev.fig} indicates, the applied corrections
appear sufficient for the data used in this study.
Therefore, we conclude that elevation effects can be successfully removed 
analytically, and we adjust all data to their zenith equivalent for all 
further analysis.

\subsection{Dependence on Time of Day}

\begin{figure}
\plotone{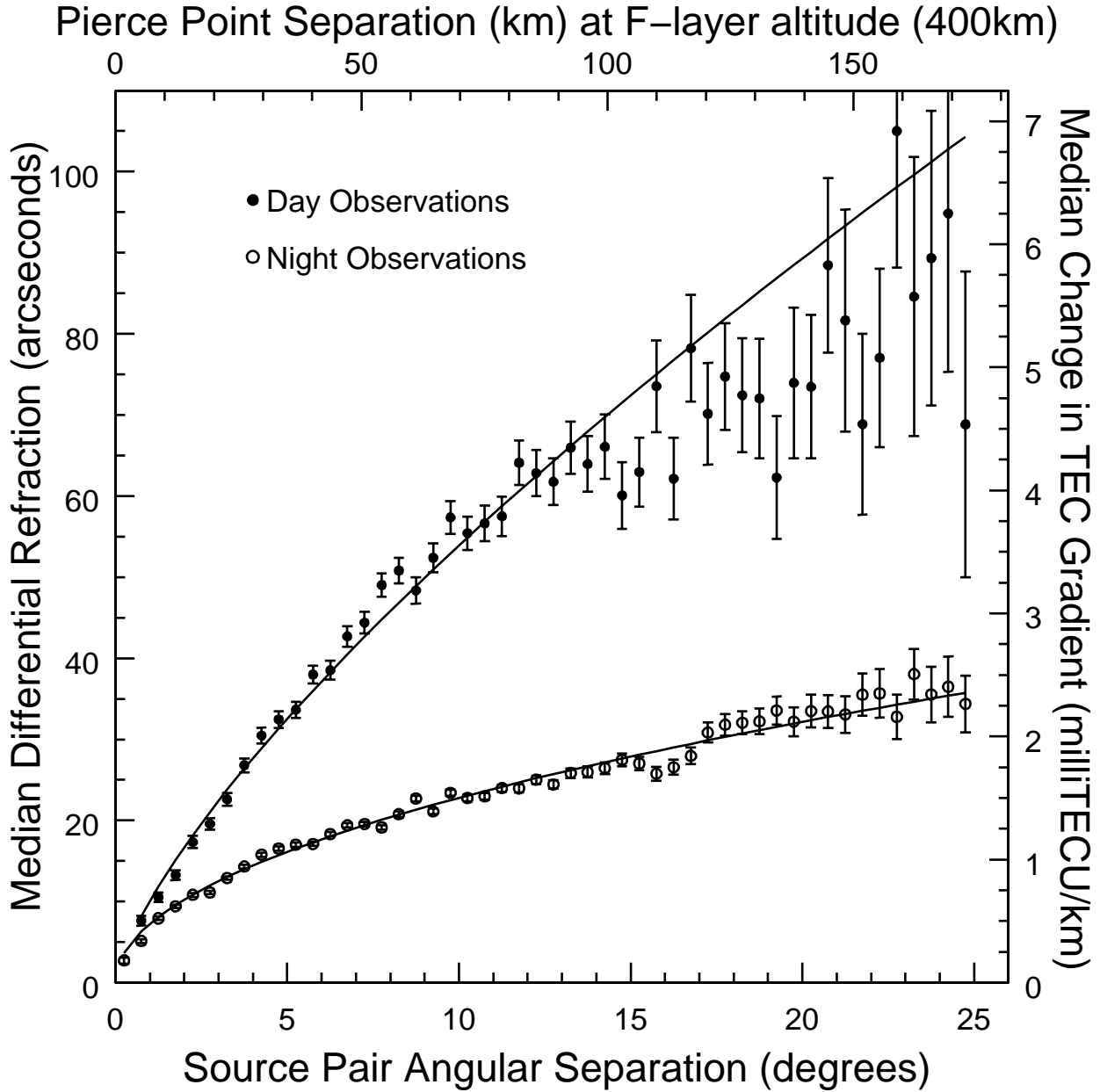}
\caption{Median differential refraction as a function of source pair separation
for data collected during the day and at night.  All data are converted to 
equivalent values at zenith.  The right axis shows the equivalent difference
in TEC gradient between the two sources as calculated using
Equation~\ref{gradient.eqn}.  The top axis shows the physical separation of 
the pierce points toward the two sources at the F-layer altitude taken to 
be 400~km.
\label{ds.daynight.fig}}
\end{figure}

Correcting for elevation effects allows the examination of other effects 
with far greater accuracy because data from a variety of elevations 
can now be combined to reduce greatly the statistical measurement errors.  
In this 
section we examine variations due to the time of day.  It is known that the
ionosphere is thicker during the day, but it is not obvious how that should
affect differential refraction.  To determine this, we have split the 
data into day (8:00 to 16:00 local solar time) and night (20:00 to 4:00)
observations.  The resulting plots of median differential refraction versus 
source pair angular separation are shown in Figure~\ref{ds.daynight.fig}. 
Clearly, the median differential refraction is 
significantly greater during the day.  For reference, on the right axis of 
this plot we have converted the differential refraction into the equivalent 
TEC gradient difference using Equation~\ref{gradient.eqn}.  Also, on the top
axis, we give the source-pair angular separation as a spatial 
separation of the source-pair pierce points projected to the ionospheric 
F-layer altitude, taken to be 400~km.

\begin{figure}
\plotone{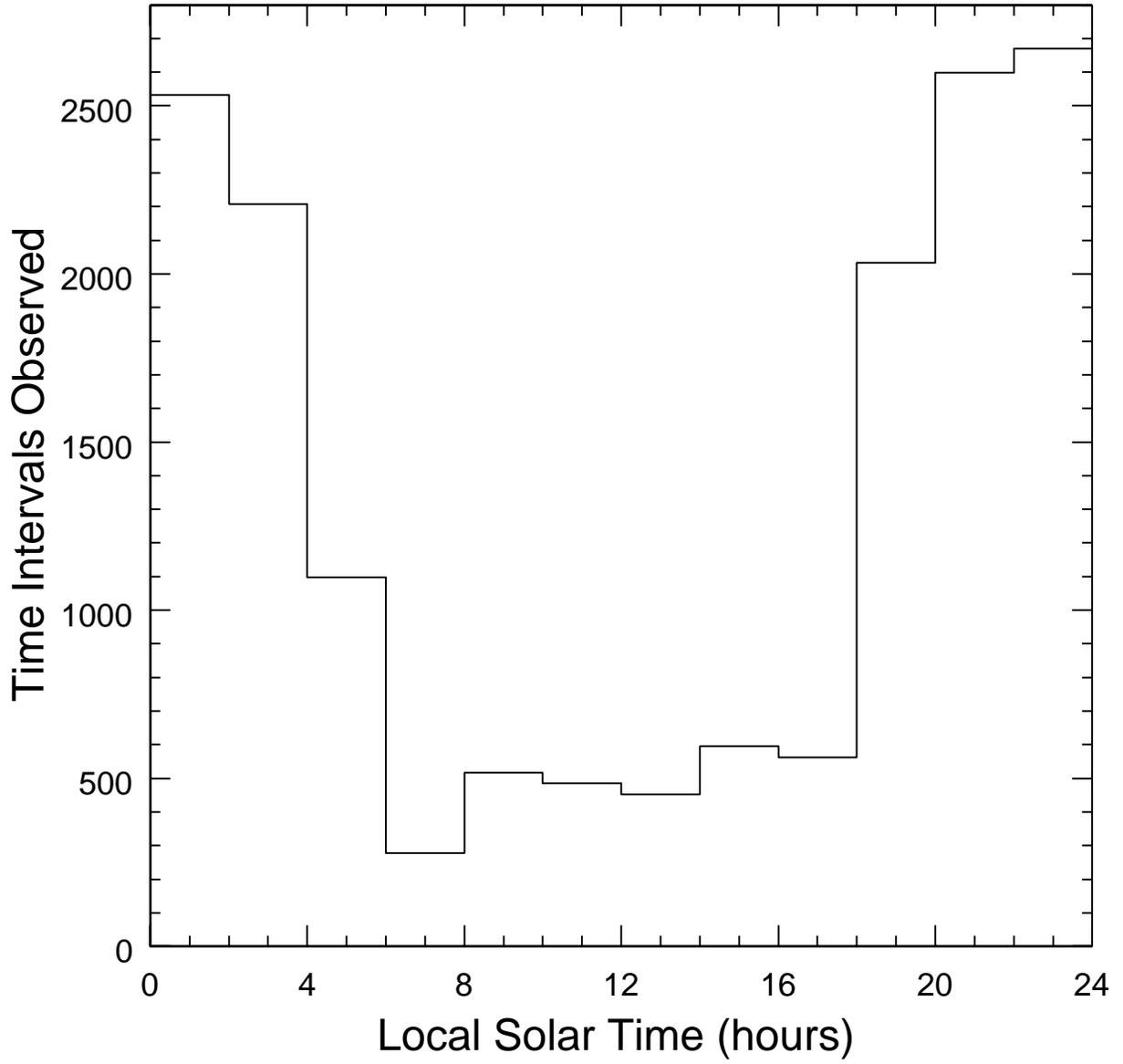}
\caption{Histogram of the number of 2-minute observations conducted as a 
function of local solar time.  Observations were done at all times, but 
mostly at night.  
\label{LST.hist.fig}}
\end{figure}

\begin{figure}
\plotone{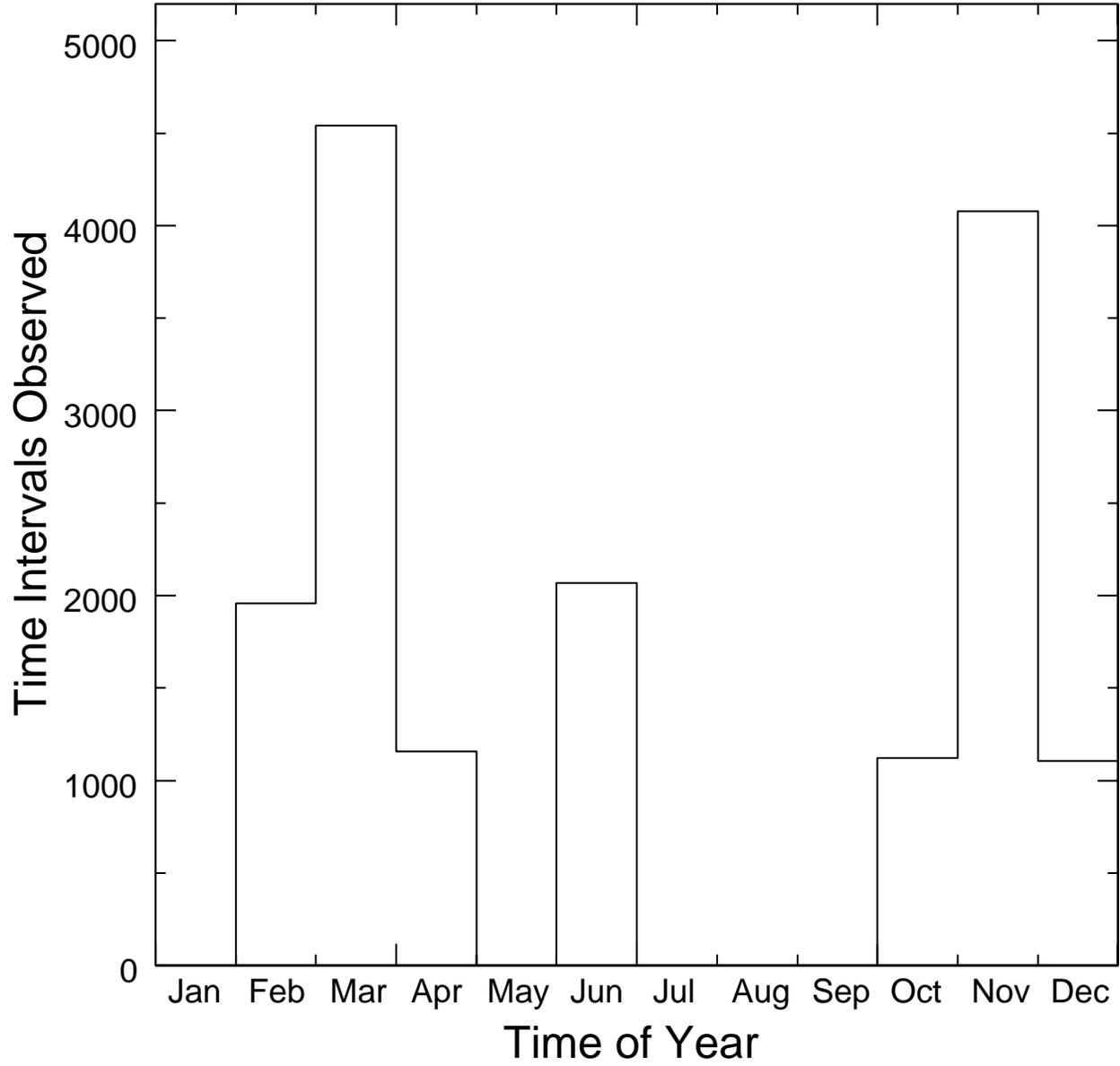}
\caption{Histogram of the number of 2-minute observations conducted as a 
function of time of year.  Because of scheduling practicalities of the VLA,
the time of year is not evenly sampled, but is biased towards the early 
spring and late fall.
\label{month.hist.fig}}
\end{figure}

The error bars generally reflect the quantity of data
available.  Less observing time was scheduled during the day than at 
night\footnote{This was done because it was already known anecdotally that 
ionospheric distortions are greater during the day.} as can be seen by the 
histogram in Figure~\ref{LST.hist.fig}.  
Also, the error bars are greater for the largest angular
separations because these sources pairs tend to be at the edges of the fields
of view where the primary antenna beam is increasingly attenuated and so 
fewer source pairs were detected.  We also note that the time of year was 
not evenly sampled, with most of our observations occurring during either 
early spring or late fall (Figure~\ref{month.hist.fig}).

We fit the data to a power law with the form given by 
Equation~\ref{powerlaw.eqn} in the same manner described in 
Section~\ref{elevation.sec}.  For night observations, the scale factor is 
$\beta = 7.26''$ and the exponent is $\alpha = 0.497$, while for the day
observations the scale factor is $\beta = 10.1''$ and the exponent is 
$\alpha = 0.728$.  These power-law fits to the data are also plotted in
Figure~\ref{ds.daynight.fig}.  Once again, the power law relation describes 
the data extremely well, with the possible exception of the largest 
separations 
during the day.  In this case, the median differential refraction does not
continue increasing quite as fast as the power law relation would predict.  
This could be caused by the fact that at such large separations, the 
ionosphere becomes completely uncorrelated and so further increases in the 
separation no longer result in increased differential refraction.

To get a better idea of how the ionospheric phase distortions vary throughout
the day, we have binned the data into 2-hour intervals of local solar time
and produced power-law fits to the data in each bin, the parameters of which 
are listed in Table~\ref{LST.fit.tab} along with fits to the day and night 
observations.  From these power-law 
fits, we predict the median differential refraction at four representative 
source pair angular separations of $2^\circ$, $4^\circ$, $8^\circ$ and 
$16^\circ$ (corresponding to F-layer spatial separations of 14, 28, 56, 
and 112~km respectively) as a function of local solar time 
(Figure~\ref{LST.all.fig}).

\section{Conclusions and Discussion}
\label{conclusions.sec}

We find that the median differential refraction of pairs of celestial radio 
sources is well described as a power-law function of the source pair angular
separation.  Though this function shows a strong elevation dependence, it 
appears to be due almost entirely to geometrical effects as was demonstrated 
by the fact that relatively simple geometrical corrections virtually 
eliminated this dependence.  The ability to ``remove'' elevation effects 
allowed us to reference all data points to their zenith-equivalent values, 
greatly increasing the amount of data used and therefore the accuracy of 
the derived statistics. 
 
We find that the power-law relation between the angular separation of 
source pairs and their median differential refraction varies greatly with 
time of day.  When comparing day and night observations, not only is the 
scale factor different but the exponent increases significantly during 
the day, from $\alpha = 0.497$ to $\alpha = 0.728$ 
(Figure~\ref{ds.daynight.fig} and Table~\ref{LST.fit.tab}).  
That the exponent is higher during the day indicates that the increase in 
ionospheric variations is proportionally greater on the larger spatial scales 
than on the smaller scales.

\begin{figure}
\plotone{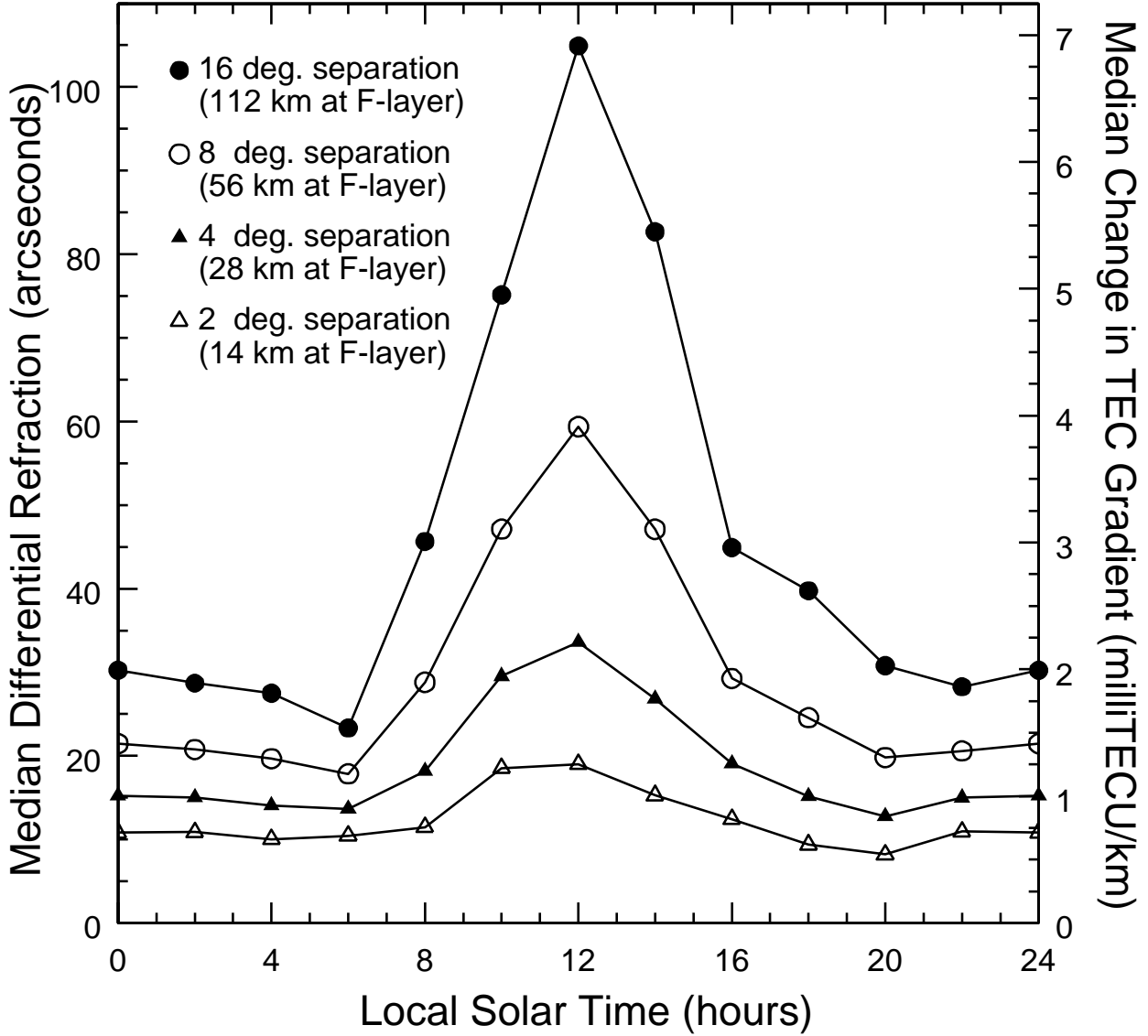}
\caption{Median differential refraction as a function of local solar time for
various source pair separations.  All data are converted to 
equivalent values at zenith and are based on power law fits to the data 
within each time range, the parameters of which are listed in 
Table~\ref{LST.fit.tab}.  
For each source pair angular separation, the 
physical separation of pierce points toward the two sources at an F-layer 
altitude of 400~km is given in parentheses.
The right axis shows the equivalent difference
in TEC gradient between the two sources as calculated using
Equation~\ref{gradient.eqn}.  
\label{LST.all.fig}}
\end{figure}

The effects of the varying exponent can be seen when examining how the 
median differential refraction varies in time for several representative 
source-pair separations as shown in Figure~\ref{LST.all.fig}.  At all 
angular separations, the median differential refraction increases during the
day in comparison to the night.  However, this increase is more dramatic
for the longer angular separations.  For example, the expected differential
refraction for a $2^\circ$ separation is about 10$''$ at midnight and  
increases by a factor of 1.5 to about 15$''$ at noon.  However, 
at a separation of $16^\circ$, the expected differential refraction increases
from $25''$ at midnight to about $80''$ at noon, which is a factor of 3.2
increase.  This demonstrates again how the additional ionospheric disturbances
seen in the daytime have a relatively greater affect on the longer 
angular spacings, indicating that they are manifested in larger-scale 
structure.

The fact that additional daytime ionospheric variations tend to come from  
predominantly large-scale structures indicates that the degree of diurnal 
variations depends heavily on the field of view as indicated in 
Figure~\ref{LST.all.fig}.  
This fact may have implications for the operation of future low-frequency 
instruments such as LOFAR, LWA and MWA.  As the diffraction-limited field of 
view is frequency dependent, it could be advantageous to perform higher 
frequency observations during the daytime, when their smaller
fields of view would be affected relatively less than the larger fields 
of view at the lower frequencies.

Finally, the magnitude of differential 
refraction at the range of angular scales we have measured can be used by 
the designers of all future low-frequency instruments to determine the 
calibration needs at various frequencies, resolutions and fields of view 
at which each instrument is designed to operate.  While we have conducted 
our measurements at 74 MHz, differential refraction varies with frequency 
in proportion to $\nu^{-2}$ as shown in Equation~\ref{gradient.eqn}.  
Therefore the measurements we have presented can easily be converted to 
predict typical ionospheric behavior at other frequencies. 

\section{Future Work}
\label{future.sec}

This paper represents just the ``tip of the iceberg'' in the use of 
low-frequency radio interferometry to characterize fine-scale ionospheric 
structure.  As explained earlier, the data used in this study did not represent
an even sampling of the time of year (Figure~\ref{month.hist.fig}) and 
time during the solar cycle.  With additional observations it should be 
possible to determine the relation between these variables and ionospheric
structure as well.  It is also planned to link radio interferometry data to 
independent ionospheric probes, such as GPS line-of-sight TEC measurements. 
Finally, these existing and future data can be modeled with various 
phenomenon such as gravity waves or turbulence to determine the 
physical causes of the observed ionospheric structure \citep{sridharan09}.

\section{Acknowledgments}

Basic research in radio astronomy at the Naval Research Laboratory is 
supported by the office of Naval Research.  The National Radio Astronomy 
Observatory is a facility of The
National Science Foundation operated under cooperative agreement by
Associated Universities, Inc.  We thank Kenneth Dymond for helpful advice.

\newpage

\begin{deluxetable}{cccr}
\tablecaption{Power-law fits for various elevation ranges.
\label{elev.fit.tab}}
\tablehead{
\colhead{Elevation Range} & 
\colhead{$\beta$} & 
\colhead{$\alpha$} & 
\colhead{$N_{Source Pairs}$}
} 
\startdata
$60^\circ$ to $90^\circ$ & 7.49 & 0.469 & 116,190 \\
$40^\circ$ to $60^\circ$ & 11.5 & 0.456 & 151,324\\
$0^\circ$ to $40^\circ$ & 20.4 & 0.444 & 31,845\\
\enddata
\tablecomments{Parameters of power-law fits (see Equation~\ref{powerlaw.eqn}) 
to the median differential refraction for various elevation ranges.
Also shown are the number of observed source pairs for each case.}
\end{deluxetable}

\begin{deluxetable}{cccr}
\tablecaption{Power-law fits for various time ranges.
\label{LST.fit.tab}}
\tablehead{
\colhead{Local Time Range} & 
\colhead{$\beta$} & 
\colhead{$\alpha$} & 
\colhead{$N_{Source Pairs}$}
}
\startdata
08:00$-$16:00 (day) & 10.1 & 0.728 & 37,388  \\
20:00$-$04:00 (night) & 7.26 & 0.497 & 179,783 \\
01:00$-$03:00 & 7.88 & 0.466 & 43,803  \\
03:00$-$05:00 & 7.19 & 0.484 & 35,157  \\
05:00$-$07:00 & 8.00 & 0.386 & 9,404  \\
07:00$-$09:00 & 7.24 & 0.664 & 6,280  \\
09:00$-$11:00 & 11.6 & 0.673 & 8,802  \\
11:00$-$13:00 & 10.8 & 0.821 & 8,017  \\
13:00$-$15:00 & 8.71 & 0.812 & 11,750  \\
15:00$-$17:00 & 8.10 & 0.618 & 11,335  \\
17:00$-$19:00 & 5.80 & 0.694 & 22,817  \\
19:00$-$21:00 & 5.29 & 0.636 & 49,156  \\
21:00$-$23:00 & 7.98 & 0.456 & 44,378  \\
23:00$-$01:00 & 7.69 & 0.494 & 48,457  \\
\enddata
\tablecomments{Parameters of power-law fits (see Equation~\ref{powerlaw.eqn})  
to the median differential refraction for various time ranges given 
in local solar time.  The first two intervals are for the day and night 
data.  Subsequently, smaller time intervals are shown throughout the day.
Also shown are the number of observed source pairs for each case.  All
data used for the fits are converted to equivalent values at zenith.}
\end{deluxetable}

\newpage

\newpage

\newpage

\newpage

\newpage

\newpage

\newpage

\newpage

\newpage

\newpage

\newpage

\newpage

\end{document}